\documentclass [12pt]{article}
\usepackage {graphicx}
\usepackage {longtable}

\begin{document}

\footnotesize{\textit{International Journal of Modern Physics B}  \textbf{20}, no. 16, 2323-2337 (2006)

\bigskip

\begin{center}
\textbf{HEAVY ELECTRONS: ELECTRON DROPLETS GENERATED BY PHOTOGALVANIC AND
PYROELECTRIC EFFECTS}
\end{center}

\bigskip

\begin{center}
VOLODYMYR KRASNOHOLOVETS,
\end{center}

\begin{center}
\textit{Institute for Basic Research, 90 East Winds Court, Palm Harbor, FL, pp. 2323-2337
34683, USA}   \\
e-mail:  \textit{v\_kras@yahoo.com}
\end{center}

\begin{center}
NICOLAI KUKHTAREV and TATIANA KUKHTAREVA
\end{center}

\begin{center}
\textit{Department of Physics, Alabama A\&M University, Huntsville, Alabama
35762, USA}  \\
e-mail:  \textit{nickoly.kukhtarev@email.aamu.edu}
\end{center}

\bigskip

\textbf{Abstract.} Electron clusters, X-rays and nanosecond radio-frequency
pulses are produced by 100 mW continuous-wave laser illuminating
ferroelectric crystal of LiNbO$_{3}$. A long-living stable electron droplet
with the size of about 100 $\mu$m has freely moved with the
velocity $\sim $0.5 cm/s in the air near the surface of the crystal
experiencing the Earth gravitational field. The microscopic model of cluster
stability, which is based on submicroscopic mechanics developed in the real
physical space, is suggested. The role of a restraining force plays the
inerton field, a substructure of the particles' matter waves, which a solitary
one can elastically withstand the Coulomb repulsion of electrons. It is
shown that electrons in the droplet are heavy electrons whose mass at least
1 million of times exceeds the rest mass of free electron. Application for
X-ray imaging and lithography is discussed.

\bigskip


\bigskip

\subsection*{1. Introduction}

Wigner crystallization [1] of electrons predicted in the 1930s
received a remarkable support in the 1970s when electron crystals/clusters
were experimentally observed on the surface of liquid helium bubbles (see
e.g. review paper [2]). This allowed further theoretical studies, for
instance, such as electron crystallization in a polarizable medium [3],
clusterization of electrons [4] and computer models of two-dimensional
electron crystals [5].

In such crystals the behavior of electrons is governed by a
competition between the Coulomb repulsion and a kind of a confinement field.
Wigner associated this confinement field with the wave nature of electrons
that obey the Schr\"odinger equation [1]. In the case of a crystal, such
field is associated with phonons [6].

In the 1990s electron clusters were generated at room temperature. It has
been found by Shoulders [7,8] and others [9-12] that electron clusters,
which embraced around 10$^{10}$ electrons and had the size from several to
tens of micrometers, could easily move at a low speed in a vacuum or/and the
air passing macroscopic distances. Based on a number of data Shoulders noted
that had been obvious one got more energy out of electron clusters of
certain experiments than one put in (their cold droplets burned through
crystals). In other words, an additional energy content is available in
charged clusters. We should also point out recent results by Santilli [13]
who jointly with co-workers has been studying strongly interacting
electrons, atoms and molecules, which at special conditions are transformed
to coupled dimers and clusters; all these species have been studied in the
framework of hadronic mechanics, a new discipline that deals with strongly
interacted and coupled entities. Such entities were called ``magnecules'' by
Santilli and, in particular, he proposed a remarkable industrial application
of his invention, namely, the magnegas technology (the combustion of
magnegas, which is produced from liquid waste, liberates heat of about 0.8
times of the natural gas at extremely clean exhaust) [13].

Thus the phenomenon of strongly coupled entities and, in particular, charged
clusters is of interest to both academic and applied studies.

In this paper we present experimental results on the self-organized
spatio-temporal pattern formation during scattering of laser light in the
photorefractive ferroelectric crystals of LiNbO$_{3}$. Some of these
patterns may be explained by assuming formation of charged clusters (or
electron droplets). Laser-induced nonlinear (holographic) scattering in
electro-optic crystal allows self-visualization of the space-charge waves,
which are formed in the crystal volume by recording of set of holographic
gratings. These holographic gratings are recorded by interference patterns
formed between pump beam and scattered waves, producing moving space-charge
waves (SCW) inside the crystal. Due to electro-optic effect these SCW
modulate refractive index (forming holographic gratings) and pump beam
diffract on these holographic gratings and thus visualizing SCW [14,15].
These results are relatively well understood for the volume SCW.

We have found unusual behavior of the specular reflection (reflection of
the laser beam from the front crystal surface), that resemble visualization
of the volume SCW. We hypothesised that observed moving scattering spots may
be due to the formation of charged plasma clusters (electron droplets)
formed near the crystal surface. This assumption is supported also by the
fact that strong radio-frequency signals (with maximum in MHz region) are
picked-up by the needle-shaped and plain electrodes with capacitive
coupling.

In the present paper we present experimental results on the generation of
electron droplets (plasma clusters) at the illumination of the LiNbO$_{3}$
crystal by a focused laser beam (CW green laser, $\lambda = 532$ nm, $P =
100$ mW). These droplets are stable and slowly move along the charged
crystal surface passing at last a few centimetres before annihilation.
Moreover, we carry out a detailed theoretical study of the confinement field
as such, which holds electrons in a cluster. Namely, based on the theory of
real physical space by Bounias and Krasnoholovets [16-19] and submicroscopic
mechanics [20-25] developed by one of the authors, we disclose the inner
reasons confining electrons together.

\subsection*{2. The electron clusters formation}

Illumination of the LiNbO$_{3}$ crystal by a focused laser beam (CW green
laser, $\lambda _{{\kern 1pt} \rm laser} = 532$ nm, $P = 100$ mW), produce in
the specular reflection spots periodically generated bright droplets that
moves along the crystal surface (Fig.1).
\begin{figure}
\begin{center}
\includegraphics[scale=0.9]{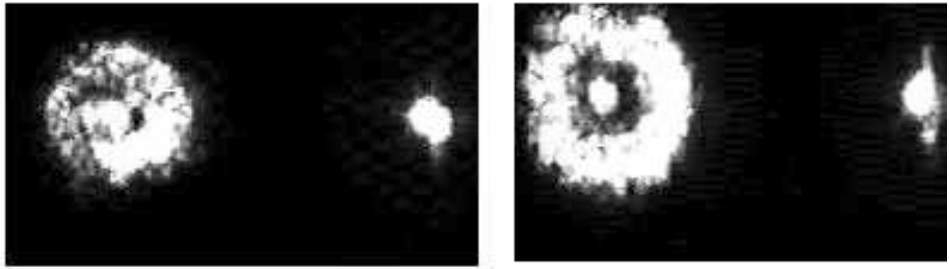}
\caption{\small{Nonlinear enhanced back-reflected scattering from the
ferroelectric crystal surface as seen on the screen with the hole (left side
of each picture). On the right side is the specular reflected beam with a
`droplet' for two frames (separated by 1 sec) from the video.}}
\label{Figure 1}
\end{center}
\end{figure}
The typical signal of the laser-initiated discharge from the crystal surface
in the air is shown on Fig. 2. In this experiment the laser beam was focused
on the Z-face of the crystal. The signal was picked up by the needle
electrode that touched the crystal surface near the illuminated spot. The
needle-type electrode enhances the electric field near the crystal surface
and facilitates plasma formation and provokes discharges. At the same time
the needle electrode served as antenna, delivering a radio-signal burst to
the oscilloscope.
\begin{figure}
\begin{center}
\includegraphics[scale=0.8]{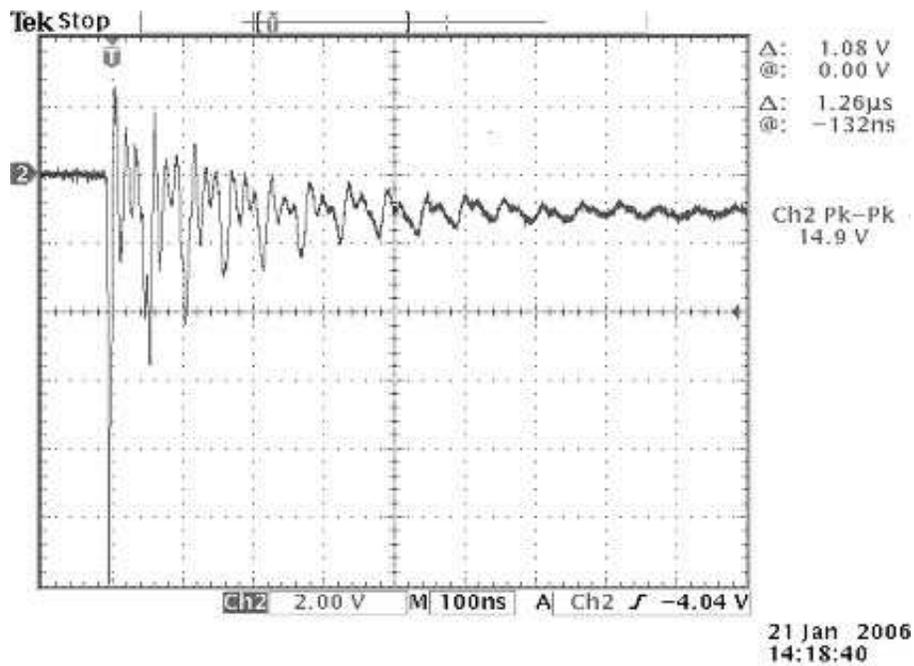}
\caption{\small{Electrical signal, picked up by the needle electrode from
the laser-illuminated spot of the LiNbO$_{3}$ crystal. Front edge of the
electrical signal is in the nanosecond scale, damping oscillations are due
to the reflections in the transmission line.}}
\label{Figure 2}
\end{center}
\end{figure}

Creation and annihilation of plasma clusters (droplets) generate nanosecond
RF pulses and (when placed in modest vacuum) bursts of X-rays. It was
already demonstrated that the intensity of X-ray generation from
ferroelectric crystals is strong enough to produce a shadow image on the
commercial dental X-ray films [27]. We suggest the explanation, based on
analogy with electron clusters (called ``EV'' by Shoulders [7,8] and
``ectons'' by Mesyats in Ref. [11,12]) in plasma discharges, created by
nonlinear photogalvanic and pyroelectric effects, that charges ferroelectric
crystals to high-voltage and initiates gas discharges. The important
difference between our results and results [7,8,11,12] is that in our case
there was no externally applied electric field; the electric field was
generated inside the crystal due to photogalvanic and pyroelectric effects
induced by the laser illumination. So it would be better to call our charged
clusters ``photogalvanic ectons'' (or ``pectons'').

Since the velocity of a relatively stable droplet is about 0.5 cm/s, which
is very small, it seems that the droplet cannot be composed of only
electrons. At the same time the droplet must be charged, because it creates
high electric field (several kV/cm) needed to modulate optical reflection
from the crystal surface [26,27]. The electron droplet is created in a
triple point, i.e. the contact of crystal and metal with the air bubbles
near the crystal surface [11,12,26]. In our case metal contact is due to
antireflection coating, and electron clouds are formed due to the photo
ionization of the crystal surface and due to the $E$-field ionization of the
air gas near the surface. Because of that, it is reasonable to assume that
electron clouds may also include positive ions from the air, though their
quantity seems rather inessential.

\subsection*{3. Theory}

Since the velocity of droplets revealed in the experiment is extremely low,
we must conjecture that electrons in a droplet are found under very peculiar
conditions, because as a matter of fact free electrons have never been
observed at a velocity lower than 10$^{6}$ m/s. However, these charged
droplets in fact have been fixed in our experiments. They moved
longitudinally to the crystal surface, which in the present case of Lithium
Niobate is characterized by a strong longitudinal electric field (several
kV/cm) perpendicular to the axis of crystal polarization.

Recent researches in sub atomic physics [17-25,6] allow us to look at the
constitution of droplets from a deeper submicroscopic viewpoint, which is
capable to shed light on the physical processes beyond the comprehension of
conventional quantum mechanics.

\subsubsection*{\textit{3.1. The behavior of particles} }

Let us discuss now a simple model of an electron droplet considering it as a
cloud of electrons tied by a confinement field. Having eluded the Coulomb
repulsion, electrons must be attracted. To understand such conditions, we
have to follow the principles of motion of canonical particles, as
submicroscopic mechanics prescribes.

The submicroscopic mechanics of canonical particles [20-25], which has been
developed in the real physical space [17-19], allows description of quantum
systems studied on the sub atomic scale. The basis for this mechanics is a
recent theory of the real space [16-19] which shows that our ordinary space
represents a mathematical lattice of primary topological balls or
superparticles, with the size of around the Planck one, $\sim $ 10$^{-35}$
m. In this case the motion of a canonical particle is associated with strong
interaction of the particle with space which results in the creation of a
cloud of elementary excitations surrounding the particle. These excitations
have been called \textit{inertons} since they appear owing to a resistance,
i.e. inertia, which a moving particle undergoes on the side of the space. As
has been argued [20-25], the amplitude $\lambda $ of spatial oscillations of
the particle appears in conventional quantum mechanics as the de Broglie
wavelength and inertons represent a substructure of the particle's matter
waves.

Thus the sub-microscopic mechanics allows one to disclose the science far
beyond the conventional quantum mechanical formalism. The amplitude $\Lambda
$ of the particle's cloud of inertons becomes implicitly apparent through
the availability of the wave $\psi $-function. Therefore, the physical
meaning of the $\psi $-function becomes completely clear: it describes the
range of space around the particle perturbed by its inertonic effect.

As follows from sub-microscopic mechanics, the amplitude of the inerton
cloud, i.e. a distance to which the front of the inerton cloud moves away
from the core of the particle, is defined by expression
\begin{equation}
\label{eq1}
\Lambda = \lambda \,c/\upsilon
\end{equation}

\noindent
where $\lambda $ is the de Broglie wave of the particle (the amplitude of
the particle), $\upsilon $ is the particle velocity and \textit{c} is the
speed of light. In submicroscopic mechanics, a particle obeys the
oscillatory motion at which the parameter $\lambda $ characterizes a section
of the particle path where the particle velocity changes from $\upsilon $ to
zero and again rises to $\upsilon $; the particle stops in the point
$\lambda /2$ due to the transfer\textit{} of its velocity and a part of its
mass to the particle's inertons. In the next half section $\lambda /2$ the
particle reabsorbs inertons, which restores the initial particle parameters,
i.e. its velocity $\upsilon $ and mass \textit{m}. Fig. 3 shows the scheme
of such motion. Furthermore, the solutions to the equations of motion show
that the motion of the particle in the tessellation space is characterized
by the two de Broglie's relationships for the particle [20-22]:
\begin{equation}
\label{eq2}
E = h\nu ,
\quad
\lambda = h/\left( {m\upsilon}  \right)
\end{equation}

\noindent
where $\nu = 1/\left( {2T} \right)$ and $1/T$ is the period of collisions
between the particle and its inerton cloud. These two relationships result
in the Schr\"odinger equation [28]. Consequently, submicroscopic mechanics,
which is constructed in the real space, can easily be reduced into the
formalism of conventional quantum mechanics developed in the phase space
(the intercoupling between two mechanics is available in Refs. [20-25]).
\begin{figure}
\begin{center}
\includegraphics[scale=0.8]{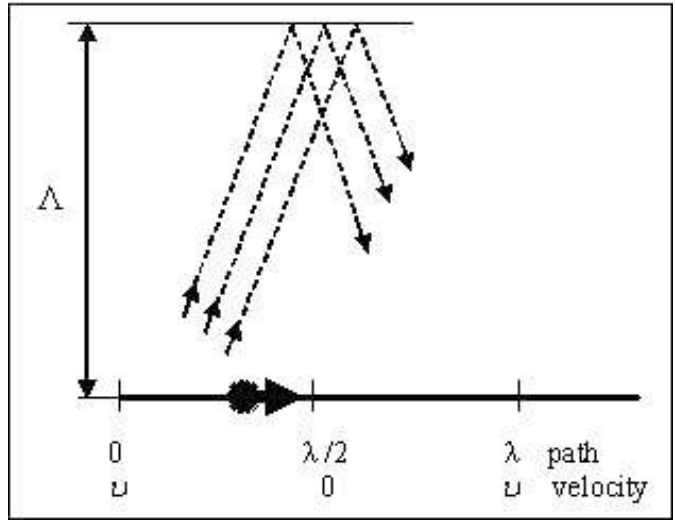}
\caption{\small{Scheme of the motion of the particle surrounded by its
cloud of inertons, as prescribed by submicroscopic mechanics.}}
\label{Figure 3}
\end{center}
\end{figure}

It is interesting to note that when the inequality $\upsilon \ll c$
holds, trajectories of inertons are practically perpendicular to the
particle path (see Fig. 3). In case of the rectilinear motion of a flow of
canonical particles, or in the case of the vibrating motion of particles,
particles' inerton clouds have to be scattered passing their energy and
momentum to the corresponding particles. If particles are not high-speeded,
the energy transfer will occur rather in directions transversal to their
path (which in the first approximation one can associate with the
\textit{X}-axis). In that case the problem can be reduced to consideration
of a flow of particles moving along the\textit{ X}-axis (or vibrating along
and against the \textit{X}-axis), which interact with each other by means of
inerton clouds along the \textit{Y}-axis. Such system of particles
interacting through their clouds of inertons can be described by the
Lagrangian
\begin{equation}
\label{eq3}
L = \sum\limits_{l} {\left( {\,{\textstyle{{1} \over {2}}} {\kern 2pt} m{\kern 1pt} \dot
{x}_{l}^{{\kern 1pt} 2} \, + \,\,\,{\textstyle{{1} \over {2}}}{\kern 2pt}\mu {\kern
1pt} {\kern 1pt} \dot {y}_{{\kern 1pt} l}^{2} \,\,\, - \,\,\omega _{{\kern
1pt} l} \sqrt {m\mu}  {\kern 1pt} {\kern 1pt} \,y_{l} {\kern 1pt} \dot
{x}_{{\kern 1pt} l} \,\, + {\kern 1pt} \,\,\tilde {\omega} _{{\kern 1pt} l}
\sqrt {m\mu}  \,{\kern 1pt} {\kern 1pt} y_{l{\kern 1pt}  +
{\kern 1pt} 1} {\kern 1pt} \dot {x}_{{\kern 1pt} l}}  \right)}
\end{equation}

\noindent
where \textit{m} and $\mu $ are masses of the \textit{l}th particle and its
inerton cloud, respectively; $x_{{\kern 1pt} l}$ and $y_{{\kern 1pt}
l{\kern 1pt}}$ are respective positions of the \textit{l}th particle and
its cloud of inertons; $y_{{\kern 1pt} l {\kern 1pt} + 1}$
is the position of the cloud of the $\left( {l + 1} \right)$th particle
(the particle is the frame of reference for its cloud of inertons);
$\omega _{{\kern 1pt} {\kern 1pt} l} = \pi /T_{{\kern 1pt} l} $
is the cyclic frequency of collisions between the \textit{l}th particle and
its inerton cloud; $\tilde {\omega} _{{\kern 1pt} {\kern 1pt} l} = \pi
/\tilde {T}_{{\kern 1pt} l}$ is the cyclic frequency of collisions
between the \textit{l} particle and the inerton cloud of
the $\left( {l + 1} \right)$th particle.

The Euler-Lagrange equations of motion for the Lagrangian (\ref{eq3}) are as follows
\begin{equation}
\label{eq4}
\ddot {x}_{l} - \sqrt {{\raise0.5ex\hbox{$\scriptstyle {\mu
}$}\kern-0.1em/\kern-0.15em\lower0.25ex\hbox{$\scriptstyle {m}$}}} {\kern
1pt} {\kern 1pt} {\kern 1pt} \omega _{{\kern 1pt} l} {\kern 2pt} \dot
{y}_{l} \,\, + \,\,\sqrt {{\raise0.5ex\hbox{$\scriptstyle {\mu
}$}\kern-0.1em/\kern-0.15em\lower0.25ex\hbox{$\scriptstyle {m}$}}} {\kern
1pt} {\kern 1pt} {\kern 1pt} \tilde {\omega} _{{\kern 1pt} l} {\kern 2pt}
\dot {y}_{l{\kern 1pt} {\kern 1pt} + {\kern 1pt} {\kern 1pt} 1} = 0;
\end{equation}
\begin{equation}
\label{eq5}
\ddot {y}_{l} + {\kern 1pt} {\kern 1pt} \sqrt
{{\raise0.5ex\hbox{$\scriptstyle
{m}$}\kern-0.1em/\kern-0.15em\lower0.25ex\hbox{$\scriptstyle {\mu} $}}}
{\kern 1pt} {\kern 1pt} {\kern 1pt} \omega _{{\kern 1pt} l} {\kern 1pt}
{\kern 1pt} \dot {x}_{l} = 0;
\end{equation}
\begin{equation}
\label{eq6}
\ddot {y}_{l{\kern 1pt} {\kern 1pt} + {\kern 1pt} {\kern 1pt} 1} {\kern 1pt}
{\kern 1pt} - {\kern 1pt} {\kern 1pt} {\kern 1pt} \sqrt
{{\raise0.5ex\hbox{$\scriptstyle
{m}$}\kern-0.1em/\kern-0.15em\lower0.25ex\hbox{$\scriptstyle {\mu} $}}}
{\kern 1pt} {\kern 1pt} \tilde {\omega} _{{\kern 1pt} l} {\kern 1pt} \dot
{x}_{{\kern 1pt} l} = 0.
\end{equation}

Since in submicroscopic mechanics the relationship $\sqrt {m/\mu}  =
c/\upsilon $ holds [19,20], the solutions to Eqs. (\ref{eq4}) to (\ref{eq6}) become
\begin{equation}
\label{eq7}
x_{l} = \frac{{\lambda _{{\kern 1pt} {\kern 1pt} l}} }{{{\kern 1pt} \pi
}}{\kern 1pt} {\kern 1pt} {\kern 1pt} \left( { - 1} \right)^{\left[ {t/T_{l}
} \right]}{\kern 1pt} {\kern 1pt} \sin\left( {\sqrt {\omega _{{\kern 1pt}
l}^{{\kern 1pt} 2} + \tilde {\omega} _{{\kern 1pt} l}^{2}}  {\kern 1pt}
{\kern 1pt} t} \right)
\end{equation}
\begin{equation}
\label{eq8}
\dot {x}_{l} = \upsilon {\kern 1pt} {\kern 1pt} {\kern 1pt} \left| {{\kern
1pt} \cos\left( {\sqrt {\omega _{{\kern 1pt} l}^{{\kern 1pt} 2} + \tilde
{\omega} _{{\kern 1pt} l}^{2}}  {\kern 1pt} {\kern 1pt} t} \right)}
\right|{\kern 1pt}
\end{equation}
\begin{equation}
\label{eq9}
\left( {y_{l} - y_{l{\kern 1pt} {\kern 1pt} + {\kern 1pt} {\kern 1pt} 1}}
\right) = \frac{{\omega _{{\kern 1pt} l}} }{{\sqrt {\omega _{{\kern 1pt}
l}^{{\kern 1pt} 2} + \tilde {\omega} _{{\kern 1pt} l}^{2}} } }{\kern 2pt} \frac{{\Lambda
_{{\kern 1pt} l}} }{{\pi} }{\kern 1pt} {\kern 1pt} {\kern 1pt} \left[
{\left( { - 1} \right)^{\left[ {t/T_{l}}  \right]} \cos\left( {\sqrt {\omega
_{{\kern 1pt} l}^{{\kern 1pt} 2} + \tilde {\omega} _{{\kern 1pt} l}^{2}}
{\kern 1pt} {\kern 1pt} t} \right) - 1} \right]
\end{equation}

The notation $\left[ {t/T_{l}}  \right]$ means an integral part of the
integer $t/T_{{\kern 1pt} l} $. Since here the parameter \textit{t}
characterizes the proper time of the \textit{l}th particle, the
corresponding coordinate (\ref{eq7}) is also treated as proper, i.e. it continuously
increases in the positive direction in both cases of the motion of
particles, rectilinear and vibratory.

In submicroscopic mechanics [20-25], in the case of a free particle
relationships
\begin{equation}
\label{eq10}
\lambda = \upsilon {\kern 1pt} {\kern 1pt} T,
\quad
\Lambda = c{\kern 1pt} {\kern 1pt} T
\end{equation}

\noindent
are held. In the present case of a flow of interacting particles the
relationships change to the following:
\begin{equation}
\label{eq11}
\lambda _{{\kern 1pt} {\kern 1pt} l} = {\kern 1pt} {\kern 1pt} {\kern 1pt}
\upsilon _{{\kern 1pt} l} \,\pi /\sqrt {\omega _{{\kern 1pt} l}^{2} + \tilde
{\omega} _{{\kern 1pt} l}^{2}}  ,
\quad
\Lambda _{{\kern 1pt} l} = {\kern 1pt} {\kern 1pt} c\,\pi /\sqrt {\omega
_{{\kern 1pt} l}^{2} + \tilde {\omega} _{{\kern 1pt} l}^{2}}  .
\end{equation}

Having obtained expressions (\ref{eq7}) to (\ref{eq9}), we use the following initial
conditions: $x_{l} \left( {0} \right) = y_{l} \left( {0} \right) = \dot
{y}_{l} \left( {0} \right) = 0$ and $\dot {x}_{{\kern 1pt} l} \left( {0}
\right) = \upsilon _{{\kern 1pt} l} $.

Solutions (\ref{eq7}) to (\ref{eq9}) show that the considered overlapping of inerton clouds
does not introduce any dispersion in the flow of particles, i.e. they
continue to move along the \textit{X}-axis with the same initial velocity
$\upsilon $, which obeys the behavior prescribed by rule (\ref{eq9}), i.e. a moving
canonical particle is characterized by the spatial amplitude $\lambda $ that
signifies a section in which the particle velocity periodically decays:
$\upsilon  \  \to $ 0  \ $ \to \upsilon  \  \to $ \ 0 \  $ \to  \upsilon $ .

Thus, submicroscopic mechanics developed in the real space makes it possible
to determine the particle's position and velocity/momentum at the same
moment of time. Submicroscopic mechanics deals with the field of inertia of
moving canonical particles. Carriers of this field, inertons, fill out a
range of space covered by the particle's $\psi $-function of conventional
quantum mechanics (the volume of this range is $\approx \lambda {\kern 1pt} {\kern
1pt} \Lambda ^{2}$). In this way, inertons representing a substructure of
matter may be interpreted as carriers of quantum mechanical interactions.
These quasi-particles of the tessellattice (the real space) transfer
momentum and energy from the quantum system under consideration to the
nearest quantum object. And also inertons, as local deformations of the real
space, transfer mass properties from the particle to its surrounding [29].

\subsubsection*{\textit{3.2. Inerton field in the crystal lattice}}

On the surface of the crystal studied, a bunch of falling photons from a
laser beam (the same in the case of a bunch of photons induced by means of
the discharge [7-12]) results in an elastic response pulse on the side of
the surface. In other words, the mechanical impact of the falling photon
bunch induces a back action of the crystal lattice. Besides, the bunch's
photons induce electron emission in the impact area of the crystal boundary.
Thus the illumination expels electrons from the crystal and also creates an
echo pulse that leaves the crystal together with knocked electrons.

As it is known in conducting and semiconducting crystals, electrons
interact with vibrating quanta of the crystal lattice. The vibratory energy
of the lattice can be written as follows,
\begin{equation}
\label{eq12}
{\begin{array}{*{20}c}
{L \, = {{1} \over
{2}}}  \,\, \sum\limits_{\vec {l}\alpha}  {{\kern 1pt} {\kern 1pt} M{\kern
1pt} \dot {\xi} _{{\kern 1pt} \vec {l}\alpha} ^{{\kern 1pt}2} - \,\,{\textstyle{{1}
\over {2}}}\,\,\sum\limits_{\vec {l}\alpha , {\kern 1pt} \vec {n}\beta} ^{\quad \ \prime}
\, V_{\alpha \beta}  \left( {\vec {l} - \vec {n}} \right)
 {\kern 1pt} \xi _{{\kern 1pt} \vec {l}\alpha}  \xi _{{\kern 1pt}
{\kern 1pt} \vec {n}{\kern 1pt} \beta}  .}   \hfill \\
{\quad \quad \;} \hfill \\
\end{array}}
\end{equation}

Here $M$ is the mass of an atom located in the $\vec {l}$th site of
the crystal lattice; $\xi _{{\kern 1pt} \vec {l}\alpha}  \ \  \left(
{\alpha = 1, \; 2, \; 3} \right)$ are three components of the $\vec
{l}$th atom displaced from the equilibrium position; $\dot {\xi} _{{\kern
1pt} \vec {l}\alpha}  $ are three components of the velocity of this atom;
$V_{{\kern 1pt} \alpha {\kern 1pt} \beta}  {\kern 1pt} {\kern 1pt} \left(
{\vec {l} - \vec {n}} \right)$ are the components of the elasticity tensor
of the crystal lattice.

It is generally recognized that such elastic interaction in the lattice is
caused by electrostatic and electromagnetic interactions between atoms.
Nevertheless, it was shown in paper [30] that the inerton component should
also be present in the total interaction of atoms in the lattice. In the
$\vec {k}$-presentation the force matrix of the crystal becomes [30]
\begin{equation}
\label{eq13}
W_{\alpha \beta}  \left( {\vec {k}} \right) = \tilde {V}_{\alpha \beta}
\left( {\vec {k}} \right){\kern 1pt} {\kern 1pt} {\kern 1pt} {\kern 1pt}
{\kern 1pt} + {\kern 1pt} {\kern 1pt} {\kern 1pt} {\kern 1pt} \tilde {\tau
}_{\alpha \beta} ^{ - 1} {\kern 1pt} \sum\limits_{{\alpha} '} {{\kern 1pt}
{\kern 1pt} \tilde {\tau} _{{\kern 1pt} {\alpha} '\beta} ^{ - 1}}  \left(
{\vec {k}} \right){\kern 1pt} {\kern 1pt}
\frac{{e_{{\alpha} '}} }{{e_{\beta} } }
\end{equation}

\noindent
where the second term describes the interaction of atoms, which is
stipulated from the overlapping of their inerton clouds. The force matrix
determines basic branches of collective vibrations of the crystal $\Omega
_{s} \left( {\vec {k}} \right)$ ($s = 1,\; 2,\; 3$), which are found
from the secular equation
\begin{equation}
\label{eq14}
||\ \Omega _{{\kern 1pt} {\kern 1pt} s}^{{\kern 1pt} 2} {\kern 1pt} \left(
{\vec {k}} \right) - W_{{\kern 1pt} \alpha {\kern 1pt} \beta}  {\kern 1pt}
\left( {\vec {k}} \right)||{\kern 1pt} {\kern 1pt} {\kern 1pt} {\kern 1pt}
{\kern 1pt} = 0.
\end{equation}

Recall, inerton clouds of atoms appear simply due to the motion of atoms,
\textit{because it is the motion of an object through the real space (the
interaction of the moving object with space), which generates the inerton
field surrounding the object}.

Coming back to the problem of generation of the echo pulse, we can now state
that it transfers not only a pulse of a classical acoustic wave, but also
includes a flow of inertons. If acoustic waves can propagate only in a
medium, a flow of inertons can spread coming through both a medium and a
vacuum (which is the real physical space). In other words, inertons of the
echo pulse can be re-absorbed by both atoms and free electrons. The
absorption of the inerton field by emitted electrons brings about
significant changes in the behavior of these electrons. Namely, the inerton
field is able to tie the electrons together. Since inertons transfer mass
(or in other words, local deformations of space [17]), the absorption of
inertons by electrons will mean the increase of electrons' mass.

In any case electrons are characterized by two kinds of interactions: the
Coulomb interaction and an elastic interaction caused by the overlapping of
electrons' inerton clouds (or wave $\psi - $functions in terms of the
conventional quantum mechanical formalism). The absorption of inertons
hardens the elastic interaction between electrons and hence suppresses their
Coulomb repulsion. In papers [4,6] the statistical behavior of interacting
particles was studied for different paired potentials and, in particular,
the possibility of clusterization of electrons was also analyzed. Let us
apply those results to the problem of electron droplet.

\subsubsection*{\textit{3.3. The droplet stability} }

The two kinds of interactions allows us to write the Hamiltonian of
interacting electrons in a droplet in the form typical for the model of
ordered particles, which is characterized by a certain nonzero order
parameter,
\begin{equation}
\label{eq15}
H\left( {n} \right) = \sum\limits_{s} {E_{s} {\kern 1pt} {\kern 1pt} n_{s}}
- {\textstyle{{1} \over {2}}}\sum\limits_{s,\,{s}^{{\kern 1pt}\prime}}
{V_{s{\kern 1pt} {s}^{{\kern 1pt}\prime}}
{\kern 1pt} n_{s} {\kern 1pt} n_{{s}^{{\kern 1pt}\prime}}}  + {\textstyle{{1} \over
{2}}}\sum\limits_{s,\,{s}^{{\kern 1pt}\prime}} {U_{s{\kern 1pt}
{s}^{{\kern 1pt}\prime}} {\kern 1pt} n_{s} {\kern
1pt} n_{{s}^{{\kern 1pt}\prime}}}.
\end{equation}

Here $E_{s} $ is the additive part of the electron energy (the kinetic
energy) in the \textit{s}th state. The main point of our approach is the
initial separation of the total potential of electron-electron interaction
into two terms: the repulsion and attraction components. So, in the
Hamiltonian (\ref{eq15}) the potential $V_{s{\kern 1pt} {s}^{{\kern 1.5pt}\prime}} $ represents the
paired energy of attraction and the potential $U_{s{\kern 1pt} {s}^{{\kern 1.5pt}\prime}} $ is
the paired energy of repulsion. The potentials take into account the
effective paired interaction between electrons located in states $s$ and
${s}^{{\kern 1.5pt}\prime}.$ The filling numbers $n_{{\kern 1pt} s} $ can have only two meanings:
1 (the \textit{s}th knot is occupied in the model lattice studied) or 0 (the
\textit{s}th knot is not occupied in the model lattice studied). The signs
before positive functions $V_{s{\kern 1pt} {s}^{{\kern 1.5pt}\prime}} $ and $U_{s{\kern 1pt}
{s}^{{\kern 1.5pt}\prime}} $ in the Hamiltonian (\ref{eq15}) directly specify proper signs of attraction
(minus) and repulsion (plus).

The statistical sum of the system under consideration
\begin{equation}
\label{eq16}
Z = \sum\limits_{\{ n\}}  {\exp\left( { - H\left( {n} \right)/k_{\rm B} T}
\right)}
\end{equation}

\noindent
can be presented in the field form [4,6]
\begin{equation}
\label{eq17}
Z = {\rm Re} \frac{{1}}{{2\pi {\kern 1pt} i}}\int {D\phi {\kern 1pt} {\kern 1pt}
\int {D\psi} }  {\kern 1pt} {\kern 1pt} \oint {d{\kern 1pt} z} {\kern 1.5pt} \exp\left[
{S\left( {\phi ,{\kern 1pt} {\kern 1pt} \psi ,{\kern 1pt} {\kern 1pt} z}
\right)} \right]
\end{equation}

\noindent
where the action
\begin{equation}
\label{eq18} {\begin{array}{*{20}c}
 {S = \sum\limits_{s} {\left\{ { - {\textstyle{{1}
\over {2}}}\sum\limits_{{s}'}
{\left( {\,\tilde {U}_{s{s}'}^{ - 1} {\kern 1pt} \phi _{s} {\kern
1pt} \phi _{{s}'} + \tilde {V}_{s{s}'}^{ - 1} {\kern 1pt} \psi
_{s} {\kern 1pt} \psi _{{s}'}}  \right)} \;\;\;\;\;\;\quad}
\right.}} \hfill \\
 {\quad \quad \quad \ \ \ \  \left. { + \ln\left| {{\kern 1pt} 1 + \frac{{1}}{{z}}
{\kern 1pt}\exp \left( - {\tilde E}_s  + \psi _s \right) \cos\phi
_{s}} \right|} \right\} + \left( {N - 1} \right)\ln z.}
\hfill \\
\end{array}}
\end{equation}

\noindent
and $D\phi $ and $D\psi $ imply the functional integration with respect to
abstract fields $\phi _{s} $ and $\psi _{s} $, respectively, and $z = \xi +
i\zeta $.
Then the action (\ref{eq18}) allows the modification [4,6]
\begin{equation}
\label{eq19} \begin{array}{l}
 S = {\textstyle{{1} \over {2}}}K \times \left\{ {\left( {a - b}
\right){\kern 1pt}{\kern 1pt} \aleph ^{{\kern 1pt} 2} - \left(
{\langle n\rangle ^{ - 1} - 1} \right) {\kern 1pt}
{\kern 1pt} \left( {\aleph ^{ - 1} + 1} \right)^{2}\exp\left( {2{\kern 1pt}
b{\kern 1pt} {\kern 1pt} \aleph}  \right)} \right. \\
\qquad  \left. { - \ln \left( {\aleph + 1} \right)} \right\} + \left( {\aleph - 1}
\right)\ln \xi \\
 \end{array}
\end{equation}

\noindent
where the following functions are introduced
\begin{equation}
\label{eq20}
a = \frac{{3}}{{k_{\rm B} T}}{\kern 1pt} {\kern 1pt} {\kern 1pt}
\int\limits_{1}^{\aleph ^{{\kern 1pt} 1/3}} {\,\,U\left( {\bar {r}{\kern
1pt} x} \right)} \,x^{2}\,d\,x,
\end{equation}
\begin{equation}
\label{eq21}
b = \frac{{3}}{{k_{\rm B} T}}{\kern 1pt} {\kern 1pt} {\kern 1pt}
\int\limits_{1}^{\aleph ^{{\kern 1pt} 1/3}} {\,V\left( {\bar {r}{\kern 1pt}
x} \right)} \,x^{2}\,d\,x.
\end{equation}

In expressions (\ref{eq19}) to (\ref{eq21}) $\langle n\rangle $ is the average filling
number of a lattice knot, $\aleph $\textit{} is the combined variable that
includes a mixture of fields $\phi _{s} $ and $\psi _{s} $, and the fugacity
$\xi = \exp\left( { - \mu /k_{\rm B} T} \right)$ where $\mu $ is the chemical
potential. The value of $\aleph $ exactly corresponds to the value of
particles in a cluster, such that $K{\kern 1pt}\aleph {\kern 1pt} = N$, where $\aleph $ is the
total quantity of particles in the system under considerations and $K$ is
the number of clusters. In other words, in our problem $N$ is the quantity
of all electrons emitted from the area of crystal irradiated by the laser
beam and $\aleph $ is the number of electrons that forms the cluster.
Besides, expressions (\ref{eq20}) and (\ref{eq21}) are written in dimensional form where $x$
is the dimensionless distance and $\bar {r}$ is the normalizing factor (for
instance, it may be an average distance between electrons in a cluster).

It is obvious that the repulsive paired potential for electrons is
\begin{equation}
\label{eq22}
U = {\kern 1pt} {\kern 1pt} {\kern 1pt} {\kern 1pt} \frac{{1}}{{4{\kern 1pt}
{\kern 1pt} \pi {\kern 1pt} \varepsilon _{{\kern 1pt} 0}} }{\kern 1pt}
{\kern 1pt} {\kern 1pt} {\kern 1pt} \frac{{e^{2}}}{{{\kern 1pt} \bar
{r}{\kern 1pt} x}}.
\end{equation}

An elastic interaction of electrons through the inerton field may be
presented in the form of a typical harmonic potential
\begin{equation}
\label{eq23}
V = {\textstyle{{1} \over {2}}}{\kern 2pt}m{\kern 1.5pt}\omega ^{2} \cdot \left( {\delta {\kern 1pt}
\bar {r}{\kern 1pt} {\kern 1pt} {\kern 1pt} x} \right)^{2}
\end{equation}

\noindent
where $m$ is the mass of an electron in the droplet, $\omega $ is the cyclic
frequency of its oscillations and $\delta  \bar {r}$ is the
amplitude of the electron displacement from its equilibrium state (note that
this amplitude is directly connected with the de Broglie wavelength of the
electron, $2{\kern 1pt} \delta \bar {r} = \lambda
$ [6]) .

Then preserving main terms in the action (\ref{eq19}) we obtain
\begin{equation}
\label{eq24}
S \approx {\textstyle{{1} \over {2}}}K \times \left( {\frac{{3{\kern 1pt}
{\kern 1pt} e^{{\kern 1pt} 2}}}{{8{\kern 1pt} {\kern 1pt} \pi {\kern 1pt} \varepsilon
_{0} {\kern 1pt} {\kern 1pt} \bar {r}{\kern 1pt} {\kern 1pt} k_{\rm B} T}}{\kern
1pt} {\kern 1pt} {\kern 1pt} \aleph ^{{\kern 1pt} 2/3}\,\, - \,\,\,\frac{{3{\kern 1pt}
{\kern 1pt} {\kern 1pt} m{\kern 1pt} {\kern 1pt} {\kern 1pt} \omega
^{2}{\kern 1pt} {\kern 1pt} \delta {\kern 1pt} \bar {r}^{{\kern 1pt} 2}{\kern 1pt}
{\kern 1pt} \aleph ^{{\kern 1pt} 5/3}}}{{k_{\rm B} T}}} \right){\kern 1pt}
 \aleph ^{{\kern 1pt} 2}.
\end{equation}

The minimum of action (\ref{eq24}) is reached at the solution of the equation
$\partial {\kern 1pt} S/\partial {\kern 1pt} {\kern 1pt} \aleph = 0$ (if the
inequality $\partial ^{2}S/\partial {\kern 1pt} {\kern 1pt} \aleph ^{2} > 0$
holds). With the approximation $\aleph \gg 1$ the corresponding solution is
\begin{equation}
\label{eq25}
\aleph \approx \frac{{20}}{{11}}{\kern 1pt} {\kern 1pt} {\kern 1pt}
\,\frac{{e^{2}/{\kern 1pt} \left( {4{\kern 1pt} \pi {\kern 1pt} \varepsilon
_{0} {\kern 1pt} {\kern 1pt} \bar {r}} \right)}}{{{\textstyle{{1} \over
{2}}}{\kern 1pt}m {\kern 1pt} \omega ^{2}{\kern 1pt} {\kern 1pt} \delta \bar
{r}^{{\kern 1pt}2}}};
\end{equation}

\noindent
in other words, the quantity of electrons involved in a droplet is
determined by the ratio of repulsive and attractive paired potentials.

\subsubsection*{\textit{3.4. Heavy electrons}}

Let us numerically estimate the quantity of electrons $\aleph $ in a
droplet. Expression (\ref{eq25}) can be rewritten as follows
\begin{equation}
\label{eq26}
\aleph \approx \frac{{20}}{{11}}{\kern 1pt} {\kern 1pt} {\kern 1pt}
\,\frac{{e^{2}/{\kern 1pt} \left( {4{\kern 1pt} \pi {\kern 1pt} \varepsilon
_{0} {\kern 1pt} {\kern 1pt} \bar {r}} \right)}}{{\hbar \omega} }
\end{equation}

\noindent
where $\hbar \omega $ is the vibration energy of an electron in the droplet.
The radius of generated droplets has been estimated as $R = 5 \times 10^{ -
5}$ m. If we set the concentration of electrons $n_{{\kern 1pt}\rm   electr}
\approx 10^{18}$ cm$^{-3}$, we can put the average distance between electrons
$\bar {r} = 10^{ - 8}$ m. Then we derive from expression (\ref{eq26})
\begin{equation}
\label{eq27}
\aleph \approx 4\,\, \times \frac{{10^{16}\;\left[ {\rm sec^{ - 1}}
\right]}}{{\omega} }.
\end{equation}

Putting $\omega = 10^{6}$ sec$^{-1}$ we obtain $\aleph \approx 4 \times
10^{10}$, which conforms with estimates of other researchers regarding the
number of electrons in a droplet. The fitting value of $\omega $ enlists the
support in our experimental results, as we have mentioned in Introduction
that radio-frequency signals, which have been picked-up, fall exactly within
the MHz region.

Such simple analysis shows that the kinetic energy of electrons in a droplet
should be very small in comparison with the kinetic energy of free electrons
and electrons in conductors and semiconductors where this energy is
proportional to $\omega \ge 10^{15}$ s$^{-1}$). However, if we equate energy
$\hbar \omega \cong 10^{ - 28}$ J (at $\omega = 10^{6}$ s$^{-1}$) and the
vibration energy ${\textstyle{{1} \over {2}}}{\kern 1pt}m{\kern 1pt}\omega ^{2}\delta {\kern
1pt} \bar {r}^{{\kern 2pt}2}$ of an electron, we will reveal that the amplitude of
oscillations of the electron near its equilibrium position will be
non-realistically large,  $\delta {\kern 1pt} \bar {r} \approx 10^{ - 5}$ m.

Thus the only resolution to fit the problem of small $\omega $ to the large
value of $\aleph $ is to accept an increase in the mass of electrons. Then
setting in equality
\begin{equation}
\label{eq28}
\hbar \omega = {\textstyle{{1} \over {2}}} {\kern 2pt} m^{\ast} \omega ^{2}{\kern 1pt}
{\kern 1pt} \delta {\kern 1pt} \bar {r}^{{\kern 2pt} 2}
\end{equation}

\noindent
the reasonable meaning $\delta {\kern 1pt} \bar {r} \approx \;(10^{ -
10}\;\,{\rm to}\;\,{\kern 1pt} 10^{ - 9})$ m, we derive for the effective mass of
an electron in the droplet $m^{\ast}  \approx 2 \times \left( {10^{ -
22}\;\,\,{\rm to}\;\,\,10^{ - 24}} \right)$ kg, which exceeds the rest mass of
electrons millions of times.

\subsection*{4. Discussion}

What is the origin for such large mass? At the moment of irradiation of the
crystal, the flow of energy of the laser beam squeezes a local area of the
crystal where a local deformation emerges. The back-reflected pulse removes
this deformation that then is distributed to emitted electrons.
Peculiarities of the droplet formation are not yet deeply understood,
however, major principles of the mechanism of increase in the electron mass
have become clear: the space deformation caused in the crystal by a
pulse/flow of photons passes into emitted electrons. This claim, as well as
the direct accordance between notions of a local deformation of space and
mass, are confirmed by our mathematical theory of the real physical space
[16-19, 29], which represents a detailed formalization of fundamental physics.

Since the droplet can be formed only in the course of non-adiabatic
processes, the time of formation of a droplet should be much less than the
relaxation time of electrons in a metal ($t \ll \tau _{\rm relax.} \ll 10^{
- 12}$ s) and the inverse Debye frequency ($t \ll {\kern 1pt} \,\,\nu
_{{\kern 1pt} \rm D}^{ - 1} \approx 10^{ - 13}\,\;{\rm to}\;\;10^{ - 12}$ s). The bond
energy of an electron in the droplet can be evaluated as composition
\begin{equation}
\label{eq29}
E_{{\kern 1pt} \rm bond} \,\, = \,\,\,\aleph {\kern 1pt} {\kern 1pt} \hbar
{\kern 1pt} {\kern 1pt} \omega {\kern 1pt} \,\, \approx \,\,4 \times 10^{ -
18} \ {\rm J}.
\end{equation}

This energy is injected in the crystal through a surface area comparable
with the size of the droplet, by the laser beam with power $P = 100$ mW
during a time of about $t \approx 10^{ - 16}$ s. This time satisfies the above
inequalities.

An electron can be knocked out by a hard ultraviolet photon with the
frequency $E_{\rm bond} /h = {\kern 1pt} {\kern 1pt} {\kern 1pt} {\kern 1pt} \nu
_{\rm ph} = 6 \times 10^{15}$ s, or $\lambda _{{\kern 1pt}\rm  ph} = 50$
nm. When knocking, the electron loses its heavy mass and the difference
$\Delta m = m^{\ast}  - m_{{\kern 1pt} 0} $ will transform to inertons that
are emitted from the electron and the electron is returned to its rest mass
$m_{0} $.

Note in passing that the idea of a variation in mass is important also in
thermodynamic processes [31]. Namely, in paper [31] the mass-energy relation
$ \pm \Delta {\kern 1pt} mc^{\kern 1pt 2}$ has been linked to the laws of
thermodynamics, which is also supported by the submicroscopic consideration.
Formalization of fundamental physics allows us easily to account for such
variations in mass, because mass is treated as a local deformation of the
tessellation space and this is a fractal volumetric deformation of a cell of
space [16-19] (whereas a fractal deformation of the surface of a cell
corresponds to the notion of the electric charge and electromagnetic
polarization [17,18,32]).

Therefore, the notion of the defect of mass $\Delta m$ is an inherent
property not only for atomic nuclei, but for\textit{} any physical system,
from quantum (for instance, electron) to macroscopic.

Since electron droplets consist of the mass substance, they are able to
behave as conventional matter droplets that are studied in the framework of
fluid, or hydrodynamic mechanics. That is, moving electron droplets can
change their shape with time, which one can see in Fig. 1, and such behavior
for droplets indeed is prescribed by the hydrodynamic laws (see, e.g. Refs. [33,34]).

\subsection*{5. Conclusion}

The research carried out in the present work shows that electron droplets
can be created at the irradiation of the polar (pyroelectric and
ferroelectric) crystals boundary by a laser beam at special conditions. A
detailed understanding of these conditions requires further experimental and
theoretical studies.

From the experimental results of our examination we may conclude that a part
of the photo-induced holographic scattering in the ferroelectric crystals
may be interpreted as scattering on the charged plasma clusters (electron
droplets). Needle-type electrodes greatly enhance near-surface plasma
formation, resembling formation of the high-energy electron clusters
described by Mesyats [11,12] as ``ectons''. But in contrast to previous
works [11,12] in our case there are no externally applied high-voltage
pulses: in our experiments high-voltage nanosecond electrical pulses were
generated in the ferroelectric crystal by the low-power CW laser
illumination due to photogalvanic and pyroelectric effects. For this reason
our hypothetical charged plasma (electron) clusters may be called
``photogalvanic electron clusters'', or ``pectons''.

The mechanism of the droplet formation proposed in this work enables a
natural description of the electron's confinement in terms of the
submicroscopic concept in which an important role is played by excitations
of the real space (inertons), which carry mass properties of physical
entities. We have argued that this is the inerton interaction between
electrons, which overcomes their Coulomb repulsion. The inerton interaction
growth also means a huge increase in mass of an electron in the droplet,
namely, up to millions of the rest mass $m_{0} $.

The submicroscopic concept allows many other applications in different
branches of fundamental and condensed matter physics and, in particular, in
applied studies dedicated to alternative sources of energy, because new
species may possess an additional energy content associated with the defect
of mass.

\subsection*{Acknowledgement}

This work in part has been done with the support of the Title 111 program
and DOE/HU Sub\#633254-HC1C060. The authors thank Dr. E. Andreev and Dr. I.
Gandzha for fruitful discussions.


\begin{thebibliography}{99}

\bibitem{1} E. Wigner, On the quantum correction for thermodynamic equilibrium,
\textit{Phys. Rev}. \textbf{40}, 749-759 (1932); On the interaction of
electrons in metals, \textit{Phys. Rev.} \textbf{46}, 1002-1011(1934).

\bibitem{2} P. S. Edelman, Levitating electrons, \textit{Uspekhi Fizicheskikh Nauk}
\textbf{130}, no. 4, 675-706 (1980); in Russian.

\bibitem{3} G. Rastelli and S. Ciuchi, Wigner crystallization in a polarizable
medium, cond-mat/0406079.

\bibitem{4} V. Krasnoholovets and B. Lev, Systems of particles with interaction and
the cluster formation in condensed matter, \textit{Condensed Matter Physics}
\textbf{6}, no. 1, 1-17 (2003).

\bibitem{5} A. V. Filinov, M. Bonitz, and Yu. E. Lozovik, Wigner Crystallization in
Mesoscopic 2D Electron Systems, \textit{Phys. Rev. Lett.} \textbf{86}, 3851
(2001).

\bibitem{6} V. Krasnoholovets, Clusterization of water molecules as deduced from
statistical mechanical approach, \textit{Central European Journal of
Physics} \textbf{2}, no. 4, 698-708 (2004).

\bibitem{7} K. R. Shoulders, Energy conversion using high charge density, U.S.
patent 5,018,180, May, 1991.

\bibitem{8} K. Shoulders and S. Shoulders, Observations of the role of charge
clusters in nuclear cluster reactions, \textit{Journal of New Energy}
\textbf{1,} no. 3, Fall (1996).

\bibitem{9} P. Beckmann, Electron clusters, \textit{Galilean Electrodynamics},
Sept./Oct. \textbf{1}, no. 5, 55-58 (1990).

\bibitem{10} R. W. Ziolkowski and M. K. Tippett, Collective effect in an electron
plasma system catalyzed by a localized electromagnetic wave, \textit{Phys.
Rev.} \textbf{A43}, no. 6, pp. 3066-3072, (1991).

\bibitem{11} G. A. Mesyats, Ecton processes at the cathode in a vacuum discharge,
\textit{Proc. XVIIth Int. Symposium on Discharges and Electrical Insulation
in Vacuum}, Berkeley, CA,\textit{} July 21-26 (1996), pp. 721-731.

\bibitem{12} I. Lisitsyn, H. Akiyama, G. A. Mesyats, Role of electron
clusters - ectons - in the breakdown of\textit{} solids dielectrics,
\textit{Physics of Plasma} \textbf{5}, no. 12, 4484- 4487 (1998).

\bibitem{13} R. M. Santilli, \textit{Foundations of Hadronic Chemistry. With
Applications to New Clean Energies and Fuels} (Kluwer Academic Publishers,
Boston-Dorderecht-London, 2001).\textit{}

\bibitem{14}N. Kukhtarev, T. Kukhtareva, M. Edwards\textbf{} B. Penn, D. Frazier,
H. Abdeldayem,\textbf{} P. P. Banerjee, T. Hudson, W. A. Friday,\textbf{
}Photoinduced Optical and Electrical High-Voltage Pulsations and Pattern
Formation in Photorefractive crystals, \textit{Journal of Nonlinear Optical
Physics and Materials} \textbf{11}, no. 4, 445-453 (2002).

\bibitem{15} M. Bayssie, J. D. Brownridge, N. Kukhtarev, T. Kukhtareva, J. C. Wang,
Generation of focused electron beam and X-rays by the doped LiNbO3 crystals,
\textit{Nuclear Instruments and Methods in Physics Research}, \textbf{B
241}, 913-918 (2005).

\bibitem{16} M. Bounias and V. Krasnoholovets. Scanning the structure of ill-known
spaces: Part 1. Founding principles about mathematical constitution of
space, \textit{Kybernetes: The International Journal of Systems and
Cybernetics} \textbf{32}, no. 7/8, 945-975 (2003) (also physics/0211096).

\bibitem{17} M. Bounias and V. Krasnoholovets, Scanning the structure of ill-known
spaces: Part 2. Principles of construction of physical space,
\textit{Kybernetes: The International Journal of Systems and Cybernetics}
\textbf{32}, no. 7/8, 976-1004 (2003) (also physics/0212004).

\bibitem{18} M. Bounias and V. Krasnoholovets, Scanning the structure of ill-known
spaces: Part 3. Distribution of topological structures at elementary and
cosmic scales, \textit{Kybernetes: The International Journal of Systems and
Cybernetics} \textbf{32}, no. 7/8, 1005-1020 (2003) (also physics/0301049).

\bibitem{19} M. Bounias and V. Krasnoholovets, The universe from nothing: A
mathematical lattice of empty sets, \textit{International Journal of
Anticipatory Computing Systems} \textbf{16}, 3-24 (2004), Ed.: D. Dubois
(also physics/0309102).

\bibitem{20} V. Krasnoholovets and D. Ivanovsky, Motion of a particle and the
vacuum, \textit{Physics Essays} \textbf{6}, no. 4, 554-563 (1993) (also
quant-ph/9910023).

\bibitem{21} V. Krasnoholovets, Motion of a relativistic particle and the vacuum,
\textit{Physics Essays} \textbf{10}, no. 3, 407-416 (1997) (also
quant-ph/9903077).

\bibitem{22} V. Krasnoholovets, On the nature of spin, inertia and gravity of a
moving canonical particle, \textit{Indian Journal of Theoretical Physics}
\textbf{48}, no. 2, 97-132 (2000) (also quant-ph/0103110).

\bibitem{23} V. Krasnoholovets, Space structure and quantum mechanics,
\textit{Spacetime \& Substance} \textbf{1}, no. 4, 172-175 (2000) (also
quant-ph/0106106).

\bibitem{24} V. Krasnoholovets, Submicroscopic deterministic quantum mechanics,
\textit{International Journal of Computing Anticipatory Systems}
\textbf{11}, 164-179 (2002), Ed.: D. Dubois (also quant-ph/0109012).

\bibitem{25} V. Krasnoholovets, On the origin of conceptual difficulties of quantum
mechanics, in \textit{Developments in Quantum Physics}, Eds.: F. Columbus
and V. Krasnoholovets (Nova Science Publishers Inc., New York, 2004), pp.
85-109 (also physics/0412152).

\bibitem{26} N. Kukhtarev, T. Kukhtareva, M. E. Edwards, J. Jones, M. Bayssie, J.
Wang, S. F. Lyuksyutov and M. A. Reagan, Smart photogalvanic running-grating
interferometer, \textit{Journal of Applied Physics} \textbf{97}, 054301
(2005).

\bibitem{27} N. Kukhtarev, T. Kukhtareva, M. Bayssie, J. Wang, J. D. Brownridge,
Generation of focused electron beam by pyroelectric and photogalvanic
crystals, \textit{Journal of Applied Physics} \textbf{96}, no. 11, 6794 -
6798, (2004).

\bibitem{28} L. de Broglie, \textit{Heisenberg's Uncertainty Relations and the
Probabilistic Interpretation of Wave Mechanics} (Mir, Moscow, 1986), pp.
34-42 (Russion translation).

\bibitem{29} V. Krasnoholovets, Gravitation as deduced from submicroscopic quantum
mechanics, hep-th/0205196.

\bibitem{30} V. Krasnoholovets and V. Byckov, Real inertons against hypothetical
gravitons. Experimental proof of the existence of inertons, \textit{Indian
Journal of Theoretical Physics} \textbf{48}, no. 1, 1-23 (2000) (also
quant-ph/0007027).

\bibitem{31} V. Krasnoholovets and J.-L. Tane, An extended interpretation of the
thermodynamic theory, including an additional energy associated with a
decrease in mass,\textbf{\textit{} }\textit{International Journal of
Simulation and Process Modelling} \textbf{2}, Nos. 1/2, 67-79 (2006).

\bibitem{32} V. Krasnoholovets, On the nature of the electric charge,
\textit{Hadronic Journal Supplement} \textbf{18}, no. 4, 425-456 (2003)
(also physics/0501132).


\bibitem{33} J. Eggers, Nonlinear dynamics and breakup of free-surface flows,
\textit{Reviews of Modern Physics}  \textbf{69}, no. 3, 865-930 (1997).

\bibitem{34} A. L. Yarin, Drop impact dynamics: splashing, spreading, receding,
bouncing\textit{...} \ \textit{Annual Review of} \textit{Fluid Mechanics}
\textbf{38},159-192 (2006).


\end{thebibliography}
\end{document}